\documentclass[fleqn,10pt]{wlpeerj}

\usepackage{verbatim}
\usepackage{hyperref}

\title{Computational bioacoustics with \\deep learning: a review and roadmap}

\author[1,2]{Dan Stowell}
\affil[1]{Department of Cognitive Science and Artificial Intelligence, Tilburg University, Tilburg, The Netherlands}
\affil[2]{Naturalis Biodiversity Center, Leiden, The Netherlands}
\corrauthor[1]{Dan Stowell}{d.stowell@tilburguniversity.edu}

\begin{abstract}
Animal vocalisations and natural soundscapes are fascinating objects of study, and contain valuable evidence about animal behaviours, populations and ecosystems. They are studied in bioacoustics and ecoacoustics, with signal processing and analysis an important component. Computational bioacoustics has accelerated in recent decades due to the growth of affordable digital sound recording devices, and to huge progress in informatics such as big data, signal processing and machine learning. Methods are inherited from the wider field of deep learning, including speech and image processing. However, the tasks, demands and data characteristics are often different from those addressed in speech or music analysis. There remain unsolved problems, and tasks for which evidence is surely present in many acoustic signals, but not yet realised. In this paper I perform a review of the state of the art in deep learning for computational bioacoustics, aiming to clarify key concepts and identify and analyse knowledge gaps. Based on this, I offer a subjective but principled roadmap for computational bioacoustics with deep learning: topics that the community should aim to address, in order to make the most of future developments in AI and informatics, and to use audio data in answering zoological and ecological questions.
\end{abstract}

\begin{document}

\flushbottom
\maketitle
\thispagestyle{empty}

\section*{Introduction}

Bioacoustics---the study of animal sound---offers a fascinating window
into animal behaviour, and also a valuable evidence source for
monitoring biodiversity \citep{Marler:2004,Laiolo:2010,Marques:2012,Brown:2017}.
Bioacoustics has long benefited from computational analysis methods
including signal processing, data mining and machine learning \citep{Towsey:2012,Ganchev::2017}.
Within machine learning, \textit{deep learning} (DL) has recently
revolutionised many computational disciplines: early innovations,
motivated by the general aims of artificial intelligence (AI) and
developed for image or text processing, have cascaded through to many
other fields \citep{Lecun:2015,Goodfellow:2016textbook}.
This includes audio domains such as automatic speech recognition and
music informatics \citep{Abe_er_2020,opensourceseparation:book}.

Computational bioacoustics is now also benefiting from the power of DL to
solve and automate problems that were previously considered
intractable.
This is both enabled and demanded by the twenty-first century data deluge: digital recording devices, data storage and sharing have become dramatically more widely available, and affordable for large-scale bioacoustic monitoring, including continuous audio capture \citep{Ranft:2004,Roe:2021,Webster:2016,Roch:2017}.
The resulting deluge of audio data means that a common bottleneck is the lack of person-time for trained analysts, heightening the importance of methods that can automate large parts of the workflow, such as machine learning.

The revolution in this field is real, but it is recent:
reviews and textbooks as recently as 2017 did not give much
emphasis to DL as a tool, even when focussing on machine learning for bioacoustics \citep{Ganchev::2017,Stowell:2017chapter}.
\citet{MercadoIII::2017} reviewed the ways in which artificial neural networks (hereafter, \textit{neural networks} or NNs) had been used by bioacoustics researchers; however, that review concerns the pre-deep-learning era of neural networks, which has some foundational aspects in common but many important differences, both conceptual and practical.

Many in bioacoustics are now grappling with deep learning, and there is much interesting work which uses and adapts DL for the specific requirements of bioacoustic analysis. Yet the field is immature and there are few reference works.
This review aims to provide an overview of the emerging field of deep learning for computational bioacoustics, reviewing the state of the art and clarifying key concepts. A particular goal of this review is to identify knowledge gaps and under-explored topics that could be addressed by research in the next few years.
Hence, after stating the survey methodology, I summarise the current state of the art, outlining standard good practice and the tasks themes addressed.
I then offer a `roadmap' of work in deep learning for computational bioacoustics, based on the survey and thematic analysis, and drawing in topics from the wider field of deep learning as well as broad topics in bioacoustics.

\section*{Survey Methodology}

Deep learning is a recent and rapidly-moving field. Although deep learning has been applied to audio tasks (such as speech, music) for more than ten years, its application in wildlife bioacoustics is recent and immature. Key innovations in acoustic deep learning include \citet{Hershey:2017} which represents the maturing of audio recognition based on convolutional neural networks (CNNs)---it introduced a dataset (\textit{AudioSet}) and a NN architecture (\textit{VGGish}) both now widely-used; and convolutional-recurrent neural network (CRNN) methods \citep{Cakir:2017}. The BirdCLEF data challenge announced ``the arrival of deep learning'' in 2016 \citep{Goeau:2016}. Hence I chose to constrain keyword-based literature searches, using both Google Scholar and Web of Science, to papers published no earlier than 2016. The query used was:

\texttt{(bioacoust* OR ecoacoust* OR vocali* OR "animal calls" OR "passive acoustic monitoring") AND ("deep learning" OR "convolutional neural network" OR "recurrent neural network") AND (animal OR bird* OR cetacean* OR insect* OR mammal*)}

With Google Scholar, this search yielded 987 entries. Many were excluded due to being off-topic, or being duplicates, reviews, abstract-only, not available in English, or unavailable. Various preprints (non-peer reviewed) were encountered (in arXiv, biorXiv and more): I did not exclude all of these, but priority was given to peer-reviewed published work.
With Web of Science, the same search query yielded 55 entries.
After merging and deduplication, this yielded 159 articles.
This sub-field is a rapidly-growing one: the set of selected articles grows from 5 in 2016 through to 63 in 2021.
The bibliography file published along with this paper lists all these articles, plus other articles added for context while writing the review.

\section*{State of the art and recent developments}

I start with a standard recipe abstracted from the literature, and the taxonomic coverage of the literature, before reviewing those topics in bioacoustic DL that have received substantial attention and are approaching maturity.
To avoid repetition, some of the more tentative or unresolved topics will be deferred to the later `roadmap' section, even when discussed in existing literature.

\subsection*{The Standard Recipe for Bioacoustic Deep Learning}

Deep learning is flexible and can be applied to many different tasks, from classification/regression through to signal enhancement and even synthesis of new data. The `workhorse' of DL is however classification, by which we mean assigning data items one or more `labels' from a fixed list (e.g.\ a list of species, individuals or call types). This the topic of many DL breakthroughs, and many other tasks have been addressed in part by using the power of DL classification---even image generation \citep{Goodfellow:2014}.
Classification is indeed the main use of DL seen in computational bioacoustics.

A typical `recipe' for bioacoustic classification using deep learning, applicable to very many of the surveyed articles from recent years, is as follows. Some of the terminology may be unfamiliar, and I will expand upon it in later sections of the review:
\begin{itemize}
\item
Use one of the well-known CNN architectures (ResNet, VGGish, Inception, MobileNet), perhaps pretrained from AudioSet. (These are conveniently available within the popular DL Python frameworks PyTorch, Keras, TensorFlow.)
\item
The input will be spectrogram data, typically divided into audio clips of fixed size such as 1 second or 10 seconds, which is done so that a `batch' of spectrograms fits easily into GPU memory. The spectrograms may be standard (linear-frequency), or mel spectrograms, or log-frequency spectrograms. The ``pixels'' in the spectrogram are magnitudes: typically these are log-transformed before use, but might not be, or alternatively transformed by per-channel energy normalisation (PCEN). There is no strong consensus on the `best' spectrogram format---it is likely a simple empirical choice based on the frequency bands of interest in your chosen task and their dynamic ranges.
\item
The list of labels to be predicted could concern species, individuals, call types, or something else.
It may be a binary (yes/no) classification task, which could be used for detecting the presence (occupancy) of some sound.
In many cases a list of species is used: modern DL can scale to many hundreds of species.
The system may be configured to predict a more detailed output such as a transcription of multiple sound events; I return to this later. 
\item
Use data augmentation to artificially make a small bioacoustic training dataset more diverse (noise mixing, time shifting, mixup).
\item
Although a standard CNN is common, CRNNs are also relatively popular, adding a recurrent layer (LSTM or GRU) after the convolutional layers, which can be achieved by creating a new network from scratch or by adding a recurrent layer to an off-the-shelf network architecture.
\item
Train your network using standard good practice in deep learning (for example: Adam optimiser, dropout, early stopping, and hyperparameter tuning) \citep{Goodfellow:2016textbook}.
\item
Following good practice, there should be separate data(sub)sets for training, validation (used for monitoring the progress of training and for selecting hyperparameters), and final testing/evaluation.
It is especially beneficial if the testing set represents not just unseen data items but novel conditions, to better estimate the true generalisability of the system \citep{Stowell:2018badchj}. However, it is still common for the training/validation/testing data to be sampled from the same pool of source data.
\item
Performance is measured using standard metrics such as accuracy, precision, recall, F-score, and/or area under the curve (AUC or AUROC). Since bioacoustic datasets are usually ``unbalanced'', having many more items of one category than another, it is common to account for this---for example by using \textit{macro-averaging}, calculating performance for each class and then taking the average of those to give equal weight to each class \citep{Mesaros:2016}.
\end{itemize}
This standard recipe will work well for many bioacoustic classification tasks, including noisy outdoor sound scenes. (Heavy rain and wind remains a problem across all analysis methods, including DL.) It can be implemented using a handful of well-known Python libraries: PyTorch/TensorFlow/Keras, librosa or another library for sound file processing, plus a data augmentation tool such as SpecAugment, audiomentations or kapre.
The creation and data augmentation of spectrograms is specific to audio domains, but the CNN architecture and training is standard across DL for images, audio, video and more, which has the benefit of being able to inherit good practice from this wider field.

Data augmentation helps with small and also with unbalanced datasets, common in bioacoustics. The commonly-used augmentation methods (time-shifting, sound mixing, noise mixing) are ``no regret'' in that it is extremely unlikely these modifications will damage the semantic content of the audio. Other modifications such as time warping, frequency shifting, frequency warping, modify sounds in ways which could alter subtle cues such as those that might distinguish individual animals or call types from one another. Hence the appropriate choice of augmentation methods is audio-specific and even animal-sound-specific.

The standard recipe does though have its limits. The use of the mel frequency scale, AudioSet pretraining, and magnitude-based spectrograms (neglecting some details of phase or temporal fine structure) all bias the process towards aspects of audio that are easily perceptible to humans, and thus may overlook some details that are important for fine high-resolution discriminations or for matching animal perception \citep{Morfi:2021}. All the common CNN architectures have small-sized convolutional filters, biasing them towards objects that are compact in the spectrogram, potentially an issue for broad-band sound events.

There are various works that pay close attention to the parameters of spectrogram generation, or argue for alternative representations such as wavelets. This engineering work can lead to improved performance in each chosen task, especially for difficult cases such as echolocation clicks. However, as has been seen in previous eras of audio analysis, these are unlikely to overturn standard use of spectrograms since the improvement rarely generalises across many tasks.
Networks using raw waveforms as input may overcome many of these concerns in future, though they require larger training datasets; pretrained raw-waveform networks may be a useful tool to look forward to in the near term.

\subsection*{Taxonomic Coverage}

Species/taxa whose vocalisations have been analysed through DL include:

\begin{itemize}
\item
Birds---the most commonly studied group, covered by at least 65 of the selected papers. Specific examples will be cited below, and some overviews can be found in the outcomes of data challenges and workshops \citep{Stowell:2018badchj,Joly::2019}.
\item
Cetaceans and other marine mammals---another very large subfield, covered by 30 papers in the selected set. Again, data challenges and workshops are devoted to these taxa \citep{Frazao::2020}.
\item
Bats \citep{MacAodha::2018,Chen::2020,Fujimori::2021,Kobayashi::2021,Zhang::2020,Zualkernan::2020,Zualkernan::2021}
\item
Terrestrial mammals (excluding bats):
including
primates \citep{Bain::2021,Dufourq::2021,Oikarinen::2019,Tzirakis::2020},
elephants \citep{Bjorck::2019},
sheep \citep{Wang::2021},
cows \citep{Jung::2021},
koalas \citep{Himawan::2018a}.

A particular subset of work focusses on mouse and rat ultrasonic vocalisations (USVs). These have been of interest particularly in laboratory mice studies, hence a vigorous subset of literature based on rodent USVs primarily recorded in laboratory conditions \citep{Coffey::2019,Fonseca::2021,Ivanenko::2020,Premoli::2021,Smith::2017,Steinfath::2021}.
\item
Anurans \citep{Colonna::2016,Dias::2021,Hassan::2017,Islam::2020,LeBien::2020,Xie::2021a,Xie::2020,Xie::2021}
\item
Insects \citep{Hibino::2021,Khalighifar::2021,Kiskin::2021,Kiskin::2020,Rigakis::2021,Sinka::2021,Steinfath::2021}
\item
Fish \citep{Guyot::2021,Ibrahim::2018,Waddell::2021}
\end{itemize}

Many works cover more than one taxon, since DL enables multi-species recognition across a large number of categories and benefits from large and diverse data. Some works sidestep taxon considerations by focusing on the overall ecosystem or soundscape level (``ecoacoustic'' approaches) \citep{Sethi::2020,Heath::2021,Fairbrass::2019}.

The balance of emphasis across taxa has multiple drivers.
Many of the above taxa are important for biodiversity and conservation monitoring (including birds, bats, insects), or for comparative linguistics and behaviour studies (songbirds, cetaceans, primates, rodents).
For some taxa, 
 acoustic communication is a rich and complex part of their behaviour, and their vocalisations have a high complexity which is amenable to signal analysis \citep{Marler:2004}.
On the other hand, progress is undeniably driven in part by practical considerations, such as the relative ease of recording terrestrial and diurnal species.
Aside from standard open science practices such as data sharing, progress in bird sound classification has been stimulated by large standardised datasets and automatic recognition challenges, notably the \textit{BirdCLEF} challenge conducted annually since 2014 \citep{birdclef2014,Joly::2019}.
This dataset- and challenge-based progress follows a pattern of work seen in many applications of machine learning.
Nonetheless, the allocation of research effort does not necessarily match up with the variety or importance of taxa---a topic I will return to.

Having summarised a standard recipe and the taxonomic coverage of the literature, I next review the themes that have received detailed attention in the literature on DL for computational bioacoustics.

\subsection*{Neural Network Architectures}

The ``architecture'' of a neural network is the general layout of the nodes and their interconnections, often arranged in sequential layers of processing \citep{Goodfellow:2016textbook}.
Early work applying NNs to animal sound made use of the basic ``multi-layer perceptron'' (MLP) architecture
\citep{Koops::2015,Houegnigan::2017,Hassan::2017,MercadoIII::2017}, with manually-designed summary features (such as syllable duration, peak frequency) as input.
However, the MLP is superseded and dramatically outperformed by CNN and (to a lesser extent) recurrent neural network (RNN) architectures, both of which can take advantage of the sequential/grid structure in raw or lightly-preprocessed data, meaning that the input to the CNN/RNN can be time series or time-frequency spectrogram data \citep{Goodfellow:2016textbook}. This change---removing the step in which the acoustic data is reduced to a small number of summary features in a manually-designed feature extraction process---keeps the input in a much higher dimensional format, allowing for much richer information to be presented.
Neural networks are highly nonlinear and can make use of subtle variation in this ``raw'' data. CNNs and RNNs apply assumptions about the sequential/grid structure of the data, allowing efficient training. For example, CNN classifiers are by design invariant to time-shift of the input data. This embodies a reasonable assumption (most sound stimuli do not change category when moved slightly later/earlier in time), and results in a CNN having many fewer free parameters than the equivalent MLP, thus being easier to train.

One early work applies a CNN to classify among 10 anuran species \citep{Colonna::2016}. In the same year, 3 of the 6 teams in the 2016 BirdCLEF challenge submitted CNN systems taking spectrograms as input, including the highest-scoring team \citep{Goeeau::2016}.
Reusing a high-performing CNN architecture from elsewhere is very popular now, but was possible even in 2016: one of the submitted systems re-used a 2012 CNN designed for images, called AlexNet.
Soon after, \citet{Salamon::2017} and \citet{Knight::2017} also found that a CNN outperformed the previous ``shallow'' paradigm of bioacoustic machine learning.

CNNs are now dominant: at least 80 of the surveyed articles made use of CNNs (sometimes in combination with other modules).
Many articles empirically compare the performance of selected NN architectures for their tasks, and configuration options such as the number of CNN layers \citep{Wang::2021,Li::2021,Zualkernan::2020}.
\citet{Oikarinen::2019} studied an interesting dual task of simultaneously inferring call type and caller ID from devices carried by pairs of marmoset monkeys, evaluating different types of output layer for this dual-task scenario.

While many of the surveyed articles used a self-designed CNN architecture, 
there is a strong move towards using, or at least evaluating, off-the-shelf CNN architectures
\citep{Lasseck:2018birdclef,Zhong::2020,Guyot::2021,Dias::2021,Li::2021,Kiskin::2021,BravoSanchez::2021,Gupta::2021}.
These are typically CNNs that have been influential in DL more widely, and are now available conveniently in DL frameworks (Table \ref{tab:otsarches}).
They can even be downloaded already pretrained on standard datasets, to be discussed further below.
The choice of CNN architecture is rarely a decision that can be made from first principles, aside from general advice that the size/complexity of the CNN should generally scale with that of the task being attempted \citep{Kaplan:2020}.
Some of the popular recent architectures (notably ResNet and DenseNet) incorporate architectural modifications to make it feasible to train very deep networks; others (MobileNet, EfficientNet, Xception) are designed for efficiency, reducing the number of computations needed to achieve a given level of accuracy \citep{Canziani:2016}.

\begin{table}[th]
    \centering
    \begin{tabular}{l|r}
    CNN architecture & Num articles \\
    \hline
ResNet \citep{he2016deep} & 23 \\
VGG \citep{simonyan2014very} or VGGish \citep{Hershey:2017} & 17 \\
DenseNet \citep{Huang:2016} & 7 \\
AlexNet \citep{krizhevsky2012imagenet} & 5 \\
Inception \citep{Szegedy:2014} & 4 \\
LeNet \citep{lecun1998gradient} & 3 \\
MobileNet \citep{sandler2018mobilenetv2} & 2 \\
EfficientNet \citep{tan2019efficientnet} & 2 \\
Xception \citep{Chollet:2017} & 2 \\
U-net \citep{Ronneberger:2015} & 2 \\
\hline
Self-designed CNN & 18 \\
    \end{tabular}
    \caption{Off-the-shelf CNN architectures used in the surveyed literature, and the number of articles using them. This is indicative only, since not all articles clearly state whether an off-the-shelf model is used, some articles use modified/derived versions, and some use multiple architectures.}
    \label{tab:otsarches}
\end{table}

The convolutional layers in a CNN layer typically correspond to non-linear filters with small ``receptive fields'' in the axes of the input data, enabling them to make use of local dependencies within spectrogram data.
However, it is widely understood that sound scenes and vocalisations can be driven by dependencies over both short and very long timescales.
This consideration about time series in general was the inspiration for the design of recurrent neural networks (RNNs), with the LSTM and GRU being popular embodiments \citep{Hochreiter:1997}: these networks have the capacity to pass information forwards (and/or backwards) arbitrarily far in time while making inferences.
Hence, RNNs have often been explored to process sound, including animal sound \citep{Xian::2016,Wang::2021,Madhusudhana::2021,Islam::2020,Garcia::2020,Ibrahim::2018}.
An RNN alone is not often found to give strong performance.
However, in around 2017 it was observed that adding an RNN layer after the convolutional layers of a CNN could give strong performance in multiple audio tasks, with an interpretation that the RNN layer(s) perform temporal integration of the information that has been preprocessed by the early layers \citep{Cakir::2017}.
This ``CRNN'' approach has since been applied variously in bioacoustics, often with good results
\citep{Himawan::2018a,Morfi:2018j,Gupta::2021,Xie::2020,Tzirakis::2020,Li::2019}.
However, CRNNs can be more computationally intensive to train than CNNs, and the added benefit is not universally clear.

In 2016 an influential audio synthesis method entitled \textit{WaveNet} showed that it was possible to model long temporal sequences using CNN layers with a special `dilated' structure, enabling many hundreds of time steps to be used as context for prediction \citep{vandenOord:2016}.
This inspired a wave of work replacing recurrent layers with 1-D temporal convolutions, sometimes called temporal CNN (TCN or TCNN) \citep{bai2018empirical}.
Note that whether applied to spectrograms or waveform data, these are 1-D (time only) convolutions, not the 2-D (time-frequency) convolutions more commonly used.
TCNs can be faster to train than RNNs, with similar or superior results.
TCNs have been used variously in bioacoustics since 2021, and this is likely to continue
\citep{Steinfath::2021,Fujimori::2021,Roch::2021,Xie::2021a,Gupta::2021,Gillings::2021,Bhatia:2021}.
\citet{Gupta::2021} compare CRNN against CNN+TCN, and also standard CNN architectures (ResNet, VGG), with CRNN the strongest method in their evaluation.

Innovations in NN architectures continue to be explored. 
\citet{Vesperini::2018} applies capsule networks for bird detection.
\citet{Gupta::2021} apply Legendre memory units, a novel type of recurrent unit, in birdsong species classification.
When we later review ``object detection'', we will encounter some custom architectures for that task.
In wider DL, especially text processing, it is popular to use a NN architectural modification referred to as ``attention'' \citep{Chorowski:2015}.
The structure of temporal sequences is highly variable, yet CNN and RNN architectures implicitly assume that the pattern of previous timesteps that are important predictors is fixed. Attention networks go beyond this by combining inputs in a weighted combination whose weights are determined on-the-fly.
(Note that this is not in any strong sense a model of auditory attention as considered in cognitive science.)
This approach was applied to spectrograms by \citet{Ren:2018attention}, and used for bird vocalisations by \citet{Morfi:2021}.
A recent trend in DL has been to use attention (as opposed to convolution or recurrence) as the fundamental building block of an NN architecture, known as ``transformer'' layers \citep{vaswani2017attention}.
Transformers are not yet widely explored in bioacoustic tasks, but given their strong performance in other domains we can expect their use to increase.
The small number of recent studies shows encouraging results \citep{Elliott:2021,Wolters:2021}.

Many studies compare NN architectures empirically, usually from a manually-chosen set of options, perhaps with evaluation over many hyperparameter settings such as the number of layers.
There are too many options to search them all exhaustively, and too little guidance on how to choose a network \textit{a priori}. \citet{Brown::2021} propose one way to escape this problem: a system to automatically construct the workflow for a given task, including NN architecture selection.

\subsection*{Acoustic Features: Spectrograms, Waveforms, and More}

In the vast majority of studies surveyed, the magnitude spectrogram is used as input data.
This is a representation in which the raw audio time series has been lightly processed to a 2D grid, whose values indicate the energy present at a particular time and frequency.
Prior to the use of DL, the spectrogram would commonly be used as the source for subsequent feature extraction such as peak frequencies, sound event durations, and more.
Using the spectrogram itself allows a DL system potentially to make use of diverse information in the spectrogram; it also means the input is a similar format to a digital image, thus taking advantage of many of the innovations and optimisations taking place in image DL.

Standard options in creating a spectrogram include the window length for the short-time Fourier transforms used (and thus the tradeoff of time- versus frequency-resolution), and the shape of the window function \citep{Jones:1995}.
Mild benefits can be obtained by careful selection of these parameters, and have been argued for in DL  \citep{Heuer::2019,Knight::2020}.
A choice more often debated is whether to use a standard spectrogram with its linear frequency axis, or to use a (pseudo-)logarithmically-spaced frequency axis such as the mel spectrogram
\citep{Xie::2019,Zualkernan::2020} or constant-Q transform (CQT) \citep{Himawan::2018a}.
The mel spectrogram uses the mel scale, originally intended as an approximation of human auditory selectivity, and thus may seem an odd choice for non-human data.
Its use likely owes a lot to convenience, but also to the fact that pitch shifts of harmonic signals correspond to linear shifts on a logarithmic scale---potentially a good match for CNNs which are designed to detect linearly-shifted features reliably.
\citet{Zualkernan::2020} even found a mel spectrogram representation useful for bat signals, with of course a modification of the frequency range.
The literature presents no consensus, with evaluations variously favouring the mel \citep{Xie::2019,Zualkernan::2020}, logarithmic \citep{Himawan::2018a,Smith::2017}, or linear scale \citep{Bergler::2019}.
There is likely no representation that will be consistently best across all tasks and taxa.
Some studies take advantage of multiple spectrogram representations of the same waveform, by ``stacking'' a set of spectrograms into a multi-channel input (processed in the same fashion as the colour channels in an RGB image) \citep{Thomas::2019,Xie::2021}.
The channels are extremely redundant with one another; this stacking allows the NN flexibly to use information aggregated across these closely-related representations, and thus gain a small informational advantage.

ML practitioners must concern themselves with how their data are normalised and preprocessed before input to a NN. Standard practice is to transform input data to have zero mean and unit variance, and for spectrograms perhaps to apply light noise-reduction such as by median filtering.
In practice, spectral magnitudes can have dramatically varying dynamic ranges, noise levels and event densities.
\citet{Lostanlen::2019,Lostanlen::2019a} give theoretical and empirical arguments for the use of \textit{per-channel energy normalisation} (PCEN), a simple adaptive normalisation algorithm.
Indeed PCEN has been deployed by other recent works, and found to permit improved performance of deep bioacoustic event detectors \citep{Allen::2021,Morfi::2021}.

As an aside, previous eras of acoustic analysis have made widespread use of mel-frequency cepstral coefficients (MFCCs), a way of compressing spectral information into a small number of standardised measurements.
MFCCs have occasionally been used in bioacoustic DL \citep{Colonna::2016,Kojima::2018,Jung::2021}. However, they are likely to be a poor match to CNN architectures since sounds are not usually considered shift-invariant along the MFCC coefficient axis.
Deep learning evaluations typically find that MFCCs are outperformed by less-preprocessed representations such as the (closely-related) mel spectrogram \citep{Zualkernan::2020,Elliott:2021}.

Other types of time-frequency representation are explored by some authors as input to DL, such as wavelets \citep{Smith::2017,Kiskin::2020}
or traces from a sinusoidal pitch tracking algorithm
\citep{Jancovic::2019}.
These can be motivated by considerations of the target signal, such as chirplets as a match to the characteristics of whale sound
\citep{glotin2017fast}.

However, the main alternative to spectrogram representations is in fact to use the raw waveform as input.
This is now facilitated by NN architectures such as WaveNet and TCN mentioned above.
DL based on raw waveforms is often found to require larger datasets for training than that based on spectrograms; one of the main attractions is to remove yet another of the manual preprocessing steps (the spectrogram transformation), allowing the DL system to extract information in the fashion needed.
A range of recent studies use TCN architectures (also called 1-dimensional CNNs) applied to raw waveform input \citep{Ibrahim::2018,Li::2019,Fujimori::2021,Roch::2021,Xie::2021a}.
\citet{Ibrahim::2018} compares an RNN against a TCN, both applied to waveforms for fish classification; 
\citet{Li::2019} applies a TCN with a final recurrent layer to bird sound waveforms.
\citet{Steinfath::2021} offer either spectrogram or waveform input for their CNN segmentation method.
\citet{Bhatia:2021} investigates bird sound synthesis using multiple methods including WaveNet.
Transformer architectures can also be applied directly to waveform data \citep{Elliott:2021}.

Some recent work has proposed trainable representations that are intermediate between raw waveform and spectrogram methods \citep{2018arXiv181205920R,Zeghidour::2021}.
These essentially act as parametric filterbanks, whose filter parameters are optimised along with the other NN layer parameters.
\citet{BravoSanchez::2021} applies a representation called SincNet, achieving competitive results on birdsong classification with a benefit of short training time.
\citet{Zeghidour::2021} apply SincNet but also introduce an alternative called LEAF, finding strong performance on a bird audio detection task.

To summarise this discussion: in many cases a spectrogram representation is appropriate for bioacoustic DL, often with (pseudo-)logarithmic frequency axis such as mel spectrogram or CQT spectrogram.
PCEN appears often to be useful for spectrogram preprocessing.
Methods using raw waveforms and adaptive front-ends are likely to gain increased prominence, especially if incorporated into some standard off-the-shelf NN architectures that are found to work well across bioacoustic tasks.

\subsection*{Classification, Detection, Clustering}

The most common tasks considered in the literature, by far, are classification and detection.
These tasks are fundamental building blocks of many workflows;
they are also the tasks that are most comprehensively addressed by the current state of the art in deep learning.

The terms \textit{classification} and \textit{detection} are used in various ways, sometimes interchangeably.
In this review I interpret `classification' as in much of ML, the prediction of one or more categorical labels such as species or call type.
Classification is very commonly investigated in bioacoustic DL.
It is most widely used for species classification---typically within a taxon family, such as in the BirdCLEF challenge \citep{Joly::2021} (see above for other taxon examples).
Other tasks studied are to classify among individual animals 
\citep{Oikarinen::2019,Ntalampiras::2021},
call types \citep{Bergler::2019a}\citep{Waddell::2021},
sex and strain (within-species) \citep{Ivanenko::2020},
or behavioural states \citep{Wang::2021}\citep{Jung::2021}.
Some work broadens the focus beyond animal sound to classify more holistic soundscape categories such as biophony, geophony, anthropophony \citep{Fairbrass::2019,Mishachandar::2021}.

\begin{figure}
    \centering
    \includegraphics[width=0.5\linewidth]{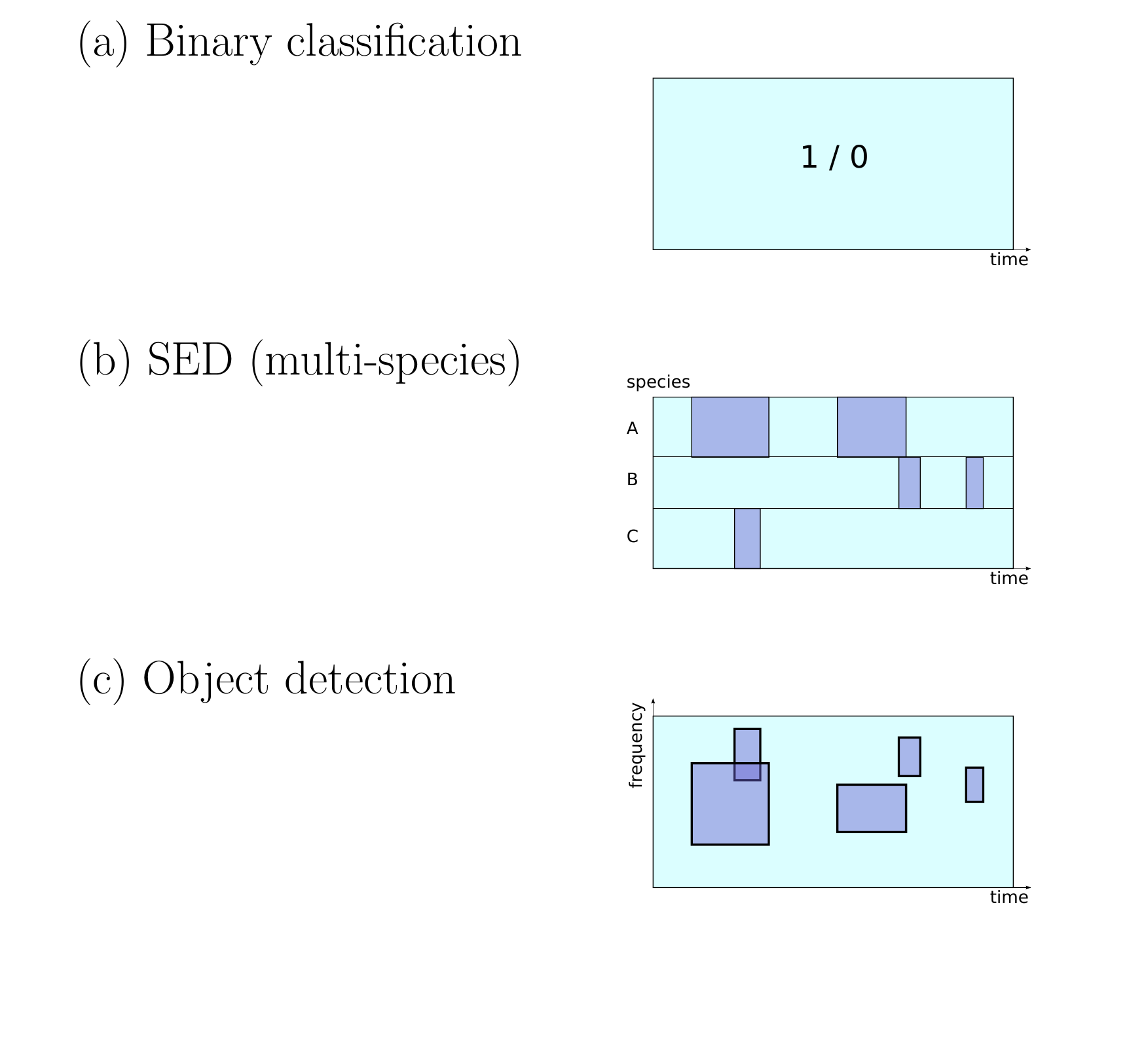}
    \caption{Three common approaches to implementation of sound detection. Adapted from \citet{Stowell:2016c}.}
    \label{fig:bdetect_paradigms}
\end{figure}

There are three different ways to define a `detection' task that are common in the surveyed literature (Figure \ref{fig:bdetect_paradigms}):
\begin{enumerate}
\item 
The first is \textbf{detection as binary classification}: for a given audio clip, return a binary yes/no decision about whether the signal of interest is detected within \citep{Stowell:2018badchj}.
This output would be described by a statistical ecologist as ``occupancy'' information.
It is simple to implement since binary classification is a fundamental task in DL, and does not require data to be labelled in high-resolution detail.
Perhaps for these reasons it is widely used in the surveyed literature
(e.g.\ \citet{MacAodha::2018,Prince::2019,Kiskin::2021,Bergler::2019,Himawan::2018a,Lostanlen::2019a}.
\item 
The second is detection as transcription, returning slightly more detail: the start and end times of sound events \citep{Morfi:2019nips4bplus,Morfi::2021}. In the DCASE series of challenges and workshops, the task of transcribing sound events, potentially for multiple classes in parallel, is termed \textbf{sound event detection (SED)}, and in the present review I will use that terminology.
It has typically been approached by training DL to label each small time step (e.g.\ a segment of 10ms or 1s) as positive or negative, and sequences of positives are afterwards merged into predicted event regions \citep{Kong:2017,Madhusudhana::2021,Marchal::2021}.
\item 
The third is the form common in \textbf{image object detection}, which consists of estimating multiple bounding boxes indicating object locations within an image. Transferred to spectrogram data, each bounding box would represent time and frequency bounds for an ``object'' (a sound event). This has not often been used in bioacoustics but may be increasing in interest \citep{Venkatesh::2021,Shrestha::2021,Zsebok::2019,Coffey::2019}.
\end{enumerate}

For all three of these task settings, CNN-based networks are found to have strong performance, outperforming other ML techniques \citep{Marchal::2021,Knight::2017,Prince::2019}.
Since the data format in each of the three task settings is different, the final (output) layers of a network take a slightly different form, as do the loss function used to optimise them \citep{Mesaros:2019}. 
(Other settings are possible, for example pixel-wise segmentation of arbitrary spectral shapes \citep{Narasimhan::2017}.)

In bioacoustics it is common to follow a two-step ``detect then classify'' workflow \citep{Waddell::2021,LeBien::2020,Schroeter::2019,Jiang::2019,Koumura::2016,Zhong::2021,Padovese::2021,Frazao::2020,Garcia::2020,Marchal::2021,Coffey::2019}.
A notable benefit of the two-step approach is that for sparsely-occurring sounds, the detection stage can be tuned to reject the large number of `negative' sound clips, with advantages for data storage/transmission, but also perhaps easing the process of training and applying the classifier, to make finer discriminations at the second step.
Combined detection and classification is also feasible, and the SED and image object detection methods imported from neighbouring disciplines often include detection and classification within one NN architecture \citep{Kong:2017,Shrestha::2021,Venkatesh::2021}.

When no labels are available even to train a classifier, unsupervised learning methods can be applied such as clustering algorithms.
The use of DL directly to drive clustering is not heavily studied.
A typical approach could be to use an unsupervised algorithm such as an autoencoder (an algorithm trained to compress and then decode data); and then to apply a standard clustering algorithm to the autoencoder-transformed representation of the data, on the assumption that this representation will be well-behaved in terms of clustering similar items together
\citep{Coffey::2019,Ozanich::2021}.

\subsection*{Signal Processing using Deep Learning}

Applications of DL have also been studied in computational bioacoustics, which do not come under the standard descriptions of classification, detection, or clustering. A theme common to the following less-studied tasks is that they relate variously to signal processing, manipulation or generation.

\textbf{Denoising and source separation} are preprocessing steps that have been used to improve the quality of a sound signal before analysis, useful in difficult SNR conditions \citep{Xie:2021}.
For automatic analysis, it is worth noting that such preprocessing steps are not always necessary or desirable, since they may remove information from the signal, and DL recognition may often work well despite noise. Denoising and source separation typically use lightweight signal processing algorithms, especially when used as a front-end for automatic recognition \citep{Xie:2021,Lin:2019}.
However, in many audio fields there is a move towards using CNN-based DL for signal enhancement and source separation \citep{opensourceseparation:book}. Commonly, this works on the spectrogram (rather than the raw audio). Instead of learning a function that maps the spectrogram onto a classification decision, denoising works by mapping the spectrogram onto a spectrogram as output, where the pixel magnitudes are altered for signal enhancement. DL methods for this are based on denoising autoencoders and/or more recently the \textit{u-net}, which is a specialised CNN architecture for mapping back to the same domain \citep{Jansson:2017}.
In bioacoustics, some work has reported good performance of DL denoising as a preprocessing step for automatic recognition, both underwater \citep{Vickers::2021,Yang::2021} and for bird sound \citep{Sinha::2018}.

\textbf{Privacy} in bioacoustic analysis is not a mainstream issue. However, Europe's GDPR regulations drive some attention to this matter, which is well-motivated as acoustic monitoring devices are deployed in larger numbers and with increased sophistication \citep{LeCornu:2021}. One strategy is to detect speech in bioacoustic recordings, in order to delete the respective recording clips, investigated for bee hive sound \citep{Janetzky:2021}. Another is to approach the task as denoising or source separation, with speech the ``noise'' to remove. \citet{Cohen:2019} take this latter approach for urban sound monitoring, and to recreate the complete anonymised acoustic scene they go one step further by blurring the speech signal content and mixing it back into the soundscape. This is perhaps more than needed for most monitoring, but may be useful if the presence of speech is salient for downstream analysis, such as investigating human-animal interactions.

\textbf{Data compression} is another concern of relevance to deployed monitoring projects. If sound is to be compressed and transmitted back for some centralised analysis, there is a question about whether audio compression codecs will impact DL analysis. \citet{Heath::2021} investigate this and concur with previous non-DL work that compression such as MP3 can have surprisingly small effect on analysis; they also obtain good performance using a CNN AudioSet embedding as a compressed `fingerprint'. \citet{Bjorck::2019} use DL more directly to optimise a codec, producing a compressed representation of elephant sounds that (unlike the fingerprint) can be decoded to the audio clip.

\textbf{Synthesis} of animal sounds receives occasional attention, and could be useful among other things for playback experimental stimuli. \citet{Bhatia:2021} studies birdsong synthesis using modern DL methods, including a WaveNet and generative adversarial network (GAN) method.

\subsection*{Small Data: Data Augmentation, Pre-training, Embeddings}

The DL revolution has been powered in part by the availability of large labelled datasets. However, a widespread and persistent issue in bioacoustic projects is the \textit{lack} of large labelled datasets: the species/calls may be rare or hard to capture, meaning not many audio examples are held; or the sound events may require a subject expert to annotate them with the correct labels (for training), and this expert time is often in short supply. Such constraints are felt for fine categorical distinctions such as those between conspecific individuals or call types, and also for large-scale monitoring in which the data volume far exceeds the person hours available. There are various strategies for effective work in such situations, including data mining and ecoacoustic methods; here I focus on techniques concerned with making DL feasible.

\textbf{Data augmentation} is a technique which artificially increases the size of a dataset (usually the training set) by taking the data samples and applying small irrelevant modifications to create additional data samples.
For audio, this can include shifting the audio in time, adding low-amplitude noise, mixing audio files together (sometimes called `mixup'), or more complicated operations such as small warpings of the time or frequency axis in a spectrogram \citep{Lasseck:2018birdclef}.
The important consideration is that the modifications should not change the meaning (the label) of the data item. Importantly, in some animal vocalisations this may exclude frequency shifts.
Data augmentation was in use even in 2016 at the ``arrival'' of DL \citep{Goeeau::2016}, and is now widespread, used in many of the papers surveyed.
Various authors study the specific combinations of data augmentation, both for terrestrial and underwater sound \citep{Lasseck:2018birdclef,Li::2021,Padovese::2021}.
Data augmentation, using the basic set of augmentations mentioned above, should be a standard part of training most bioacoustic DL systems.
Software packages are available to implement audio data augmentation directly (for example \textit{SpecAugment}, \textit{kapre} or \textit{audiomentions} for Python).
Beyond standard practice, data augmentation can even be used to estimate the impact of confounding factors in datasets \citep{Stowell:2019bgaugj}.

A second widespread technique is \textbf{pretraining}: instead of training for some task by starting from a random initialisation of the NN, one starts from a NN that has previously been trained for some other, preferably related, task. The principle of ``transfer learning'' embodied here is that the two tasks will have some common aspects---such as the patterns of time-frequency correlations in a spectrogram, which at their most basic may have similar tendencies across many datasets---and that a NN can benefit from inheriting some of this knowledge gained from other tasks.
This becomes particularly useful when large well-annotated datasets can be used for pretraining.
Early work used pretraining from \textit{image} datasets such as ImageNet, which gave substantial performance improvements even though images are quite different from spectrograms \citep{Lasseck:2018birdclef}.
Although ImageNet pretraining is still occasionally used \citep{Disabato::2021,Fonseca::2021}, many authors now pretrain using Google's AudioSet (a diverse dataset of audio from YouTube videos \citep{Hershey:2017}) \citep{Coban::2020,Kahl::2021}.
A similar but more recent dataset is VGG-Sound \citep{Chen::2020a}, used by \citet{Bain::2021}.
Practically, off-the-shelf networks with these well-known datasets are widely available in standard toolkits.
Although publicly-available bioacoustics-specific datasets (such as that from BirdCLEF) are now large, they are rarely explored as a source of pretraining---perhaps because they are not as diverse as AudioSet/VGGish, or perhaps as a matter of convenience.
\citet{Ntalampiras::2018} explored transfer learning from a music genre dataset.
Contrary to the experiences of others, \citet{Morgan::2021} report that pretraining was not of benefit in their task, perhaps because the dataset was large enough in itself (150 hours annotated).
Another alternative is to pretrain from simulated sound data, such as synthetic underwater clicks or chirps \citep{glotin2017fast,Yang::2021}.

Closely related to pretraining is the popular and important concept of \textbf{embeddings}, and (related) \textbf{metric learning}.
The common use of this can be stated simply:
instead of using standard acoustic features as input, and training a NN directly to predict our labels of interest,
we train a NN to convert the acoustic features into some partially-digested vector coordinates, such that this new representation is useful for classification or other tasks.
The ``embedding'' is the space of these coordinates.

The simplest way to create an embedding is to take a pretrained network and remove the ``head'', the final classification layers.
The output from the ``body'' is a representation intermediate between the acoustic input and the highly-reduced semantic output from the head, and thus can often be a useful high-dimensional feature representation.
This has been explored in bioacoustics and ecoacoustics using AudioSet embeddings, and found useful for diverse tasks \citep{Sethi:2020c,Sethi::2020,Coban::2020,Heath::2021}.

An alternative approach is to train an autoencoder directly to encode and decode items in a dataset, and then use the autoencoder's learnt representation (from its encoder) as an embedding \citep{Ozanich::2021,Rowe::2021}. This approach can be applied even to unlabeled data, though it may not be clear how to ensure this encodes semantic information. It can be used as unsupervised analysis to be followed by clustering \citep{Ozanich::2021}.

A third strategy for DL embedding is the use of so-called Siamese networks and triplet networks.
These are not really a separate class of network architectures---typically a standard CNN is the core of the NN.
The important change is the loss function: unlike most other tasks, training is not based on whether the network can correctly label a single item, but on the vector coordinates produced for a pair/triplet of items, and their distances from one another.
In Siamese networks, training proceeds pairwise, with some pairs intended to be close together (e.g.\ same class) or far apart (e.g.\ different class).
In triplet networks, training uses triplets with one selected as the `anchor', one positive instance to be brought close, and one negative instance to be kept far away.
In all cases, each of the items is projected through the NN independently, before the comparison is made.
The product of such a procedure is this NN trained directly to produce an embedding in which location, or at least distance, carries semantic information.
These and other embeddings can be used for downstream tasks by applying simple classification/clustering/regression algorithms to the learnt representation.
A claimed benefit of Siamese/triplet networks is that they can train relatively well with small or unbalanced datasets, and this has been reported to be the case in terrestrial and underwater projects \citep{Thakur::2019,Nanni::2020,Acconcjaioco::2021,Zhong::2021}.


Other strategies to counter data scarcity have been investigated for biacoustics:
\begin{itemize}
\item 
multi-task learning---another form of transfer learning, this involves training on multiple tasks simultaneously \citep{Morfi:2018multitask,Zeghidour::2021,Cramer::2020};
\item 
semi-supervised learning, which supplements labelled data with unlabelled data \citep{Zhong::2020b,Bergler::2019a};
\item 
weakly-supervised learning, which allows for labelling that is imprecise or lacks detail (e.g.\ lacks start and end time of sound events) \citep{Kong:2017,Knight::2017,Morfi:2018j,LeBien::2020};
\item 
self-supervised learning, which uses some aspect of the data itself as a substitute for supervised labelling \citep{Saeed:2021};
\item 
few-shot learning, in which a system is trained across multiple similar tasks, in such a way that for a new unseen task (e.g.\ a new type of call to be detected) the system can perform well even with only one or very few examples of the new task \citep{Morfi::2021,Acconcjaioco::2021}.
A popular method for few-shot learning is to create embeddings using prototypical networks, which involve a customised loss function that aims to create an embedding having good ``prototypes'' (cluster centroids). \citet{Pons:2018} determined this to outperform transfer learning for small-data scenarios, and it is the baseline considered in a recent few-shot learning bioacoustic challenge \citep{Morfi::2021}.
\end{itemize}
In general, these approaches are less commonly studied, and many authors in bioacoustics use off-the-shelf pretrained embeddings.
However, many of the above techniques are useful to enable training despite dataset limitations; hence, they can themselves be used in creating embeddings, and could be part of future work on creating high-quality embeddings.

\subsection*{Generalisation and Domain Shift}

Concern about whether the a DL system's quality of performance will generalise to new data is a widespread concern, especially when small datasets are involved.
A more specific concern is whether performance will generalise to new conditions \textit{in which attributes of the input data have changed}: for example changes in the background soundscape, the sub-population of a species, the occurrence frequency of certain events, or the type of microphone used. All of these can change the overall distribution of basic acoustic attributes, so-called \textit{domain shift}, which can have undesirable impacts on the outputs of data-driven inference \citep{Morgan::2021}.

It is increasingly common to evaluate DL systems, not only on a test set which is kept separate from the training data, but also on test set(s) which differ in some respects from the training data, such as location, SNR, or season \citep{Shiu::2020,Vickers::2021,Coban::2020,Allen::2021,Khalighifar::2021}.
This helps to avoid the risk of overestimating generalisation performance in practice.

Specific DL methods can be used explicitly to account for domain shift.
\textit{Domain adaptation} methods may automatically adapt the NN parameters \citep{Adavanne:2017,Best::2020}.
Explicitly including contextual correlates as input to the NN is an alternative strategy for automatic adaptation \citep{Lostanlen::2019a,Roch::2021}.
Where a small amount of human input about the new domain is possible, fine-tuning (limited retraining) or active learning (interactive feedback on predictions) have been explored \citep{Coban::2020,Allen::2021,Ryazanov::2021}.
\citet{Stowell:2018badchj} designed a public ``bird audio detection'' challenge specifically to stimulate the development cross-condition (cross-dataset) generalisable methods.
In that challenge, however, the leading submissions did not employ explicit domain adaptation, instead relying on the implicit generality of transfer learning (pretraining) from general-purpose datasets, as well as data augmentation to simulate diverse conditions during training.

\subsection*{Open-set and Novelty}

One problem with the standard recipe (and in fact many ML methods) is that by default, recognition is limited to a pre-specified and fixed set of labels. When recording in the wild, it is surely possible to encounter species or individuals not accounted for in the training set, which should be identified. This is common e.g. for individual ID \citep{Ptacek:2016}. 

Detecting new sound types beyond the known set of target classes is referred to as \textit{open set} recognition, perhaps related to the more general topic of novelty detection which aims to detect any novel occurrence in data.
\citet{Cramer::2020} argue that hierarchical classification is useful for this, in that a sound may be strongly classified to a higher-level taxon even when the lower-level class is novel.
\citet{Ntalampiras::2021} applies novelty detection based on a CNN autoencoder (an algorithm trained to compress and then decode data). Since the method is trained to reconstruct the training examples with low error, the authors use the assumption that novel sounds will be reconstructed with high error, and thus use this as a trigger for detecting novelty.

More broadly, the aforementioned topic of embeddings offers a useful route to handling open-set classification.
A good embedding should provide a somewhat semantic representation even of new data, such that even novel classes will cluster well in the space (standard clustering algorithms such as k-nearest neighbours can be applied). This is advocated by \citet{Thakur::2019}, using triplet learning, and later \citet{Acconcjaioco::2021} using Siamese learning.
Novelty and open-set issues are likely to be an ongoing concern, in practice if not in theory, though the increasing popularity of general-purpose embeddings indeed offers part of the solution.

\subsection*{Context and Auxiliary Information}

Deep learning implementations almost universally operate on segments of audio or spectrogram (e.g.\ 1 or 10 seconds per datum) rather than a continuous data stream. This is true even for RNNs which in theory can have unbounded time horizons. Yet it is clear from basic considerations that animal vocalisations, and their accurate recognition, may depend strongly on contextual factors originating outside a short window of temporal attention, whether this be prior soundscape activity or correlates such as date/time, location or weather.

\citet{Lostanlen::2019a} add a ``context-adaptive neural network'' layer to their CNN, whose weights are dynamically adapted at prediction time by an auxiliary network taking long-term summary statistics of spectrotemporal features as input. Similarly, \citet{Roch::2021} input acoustic context to their CNN based on estimates of the local signal-to-noise ratio (SNR).
\citet{Madhusudhana::2021} apply a CNN (DenseNet) to acoustic data, and then postprocess the predictions of that system using an RNN, to incorporate longer-term temporal context into the final output.
Note that this CNN and RNN is not an integrated CRNN but two separate stages, with the consequence that the RNN can be applied over differing (longer) timescales than the CNN.

Animal taxonomy is another form of contextual information which may help to inform or constrain inferences. Although taxonomy is rarely the strongest determinant of vocal repertoire, it may offer a partial guide. Hierarchical classification is used in many fields, including bioacoustics; \citet{Cramer::2020} propose a method that explicitly encodes taxonomic relationships between classes into the training of a CNN, evaluated using bird calls and song.
\citet{Nolasco::2021} propose a different method, and evaluate across a broader hierarchy, covering multiple taxa at the top level and individual animal identity at the lowest level.

\subsection*{Perception}

The work so far discussed uses DL as a practical tool. Deep learning methods are loosely inspired by ideas from natural perception and cognition \citep{Lecun:2015}, but there is no strong assumption that bioacoustic DL implements the same processes as natural hearing. Further, since current DL models are hard to interpret, it would be hard to validate whether or not that assumption held.

Even so, a small stream of research aims to use deep learning to model animal acoustic perception. DL can model highly non-linear phenomena, so perhaps could replicate many of the subtleties of natural hearing, which simpler signal processing models do not.
Such models could then be studied or used as a proxy for animal judgment.
\citet{Morfi:2021} use triplet loss to train a CNN to produce the same decisions as birds in a two-alternative forced-choice experiment.
\citet{Simon::2021} train a CNN from sets of bat echolocation call reflections, to classify flowers as bat-pollinated or otherwise---a simplified version of an object recognition task that a nectarivorous/frugivorous bat presumably solves.
\citet{Francl:2020} study sound localisation, finding that a DL trained to localise sounds in a (virtual) reverberant environment exhibits some phenomena known from human acoustic perception.

\subsection*{On-device Deep Learning}

Multiple studies focus on how to run bioacoustic DL on a small hardware device, for affordable/flexible monitoring in the field.
Many projects do not need DL running in real-time on device: they can record audio to storage or transmit it to a base station, for later processing \citep{Roe:2021,Heath::2021}. However, implementing DL on-device allows for live readouts and rapid responses \citep{MacAodha::2018}, potential savings in power or data transmission costs, and enables some patterns of deployment that might not otherwise be possible.
One benefit of wide interest might be to perform a first step of detection/filtering and discard many hours of uninformative audio, to extend deployment durations before storage is full and reduce transmission bandwidths: this is traditionally performed with simple energy detection, but could be enhanced by lightweight ML algorithms, perhaps similar to ``keyword spotting'' in domestic devices \citep{Zhang:2017}.

The Raspberry Pi is a popular small Linux device, and although low-power it can have much of the functionality of a desktop computer, such as running Python or R scripts; thus the Raspberry Pi has been used for acoustic monitoring and other deployments  \citep{Jolles:2021}. Similar devices are Jetson Nano and Google Coral (the latter carries a TPU unit on board intended for DL processing). \citet{Zualkernan::2021} evaluate these three for running a bat detection algorithm on-device.

Even more constrained devices offer lower power consumption (important for remote deployment powered by battery or solar power), lower ecological footprint, and smaller form factor; often based on the ARM Cortex-M family of processors. The AudioMoth is a popular example \citep{Hill:2017}. It is too limited to run many DL algorithms; however \citet{Prince::2019} were able to implement a CNN (depthwise-separable to reduce the complexity), applied to mel frequency features, and report that it outperformed a HMM detector on-device, although ``at the cost of both size and speed'': it was not efficient enough to run in real-time on AudioMoth.
Programming frameworks help to make such low-level implementations possible: \citet{Disabato::2021} use ARM CMSIS to implement a bird detector, and \citet{Zualkernan::2021} use TensorFlow Lite to implement a bat species classifier.
As in the more general case, off-the-shelf NN architectures can be useful, including MobileNet and SqueezeNet which are designed to be small/efficient \citep{VidanaVila:2020}. However, all three of the bioacoustic studies just mentioned, while inspired by these, implemented their own CNN designs and feature modifications to shrink the footprint even further.

Small-footprint device implementations offer the prospect of DL with reduced demands for power, bandwidth and storage. However, \citet{Lostanlen:2021} argue that energy efficiency is not enough, and that fundamental resource requirements such as the rare minerals required for batteries are a constraint on wider use of computational bioacoustic monitoring. They propose batteryless acoustic sensing, using novel devices capable of intermittent computing whenever power becomes available. It remains to be seen whether this intriguing proposal can be brought together with the analytical power of DL.

\subsection*{Workflows and Other Practicalities}

As DL comes into increased use in practice, questions shift from the proof-of-concept to the integration into broader workflows (e.g.\ of biodiversity monitoring), and other practicalities. Many of the issues discussed above arise from such considerations.
Various authors offer recommendations and advice for ecologists using DL \citep{Knight::2017,Rumelt::2021,Maegawa::2021}.
Others investigate integration of a CNN detector/classifier into an overall workflow including data acquisition, selection and labeling \citep{LeBien::2020,Morgan::2021,Ruff::2021}.
\citet{Brown::2021} go further and investigate the automation of designing the overall workflow, arguing that ``[t]here is merit to searching for workflows rather than blindly using workflows from literature. In almost all cases, workflows selected by [their proposed] search algorithms (even random search, given enough iterations) outperformed those based on existing literature.''

One aspect of workflow is the user interface (UI) through which an algorithm is configured and applied, and its outputs explored.
Many DL researchers provide their algorithms as Python scripts or suchlike, a format which is accessible by some but not by all potential users.
Various authors provide GUI interfaces for the algorithms they publish, and to varying extents study efficient graphical interaction
\citep{Jiang::2019,Coffey::2019,Steinfath::2021,Ruff::2021}.


\section*{A Roadmap for Bioacoustic Deep Learning}

I next turn to the selection of topics that are unresolved and/or worthy of further development: recommended areas of focus in the medium-term for research in deep learning applied within computational bioacoustics.
The gaps are identified and confirmed through the literature survey, although there will always be a degree of subjectivity in the thematic synthesis.

Let us begin with some principles.
Firstly, AI does not replace expertise, even though this may be implied by the standard recipe and general approach (i.e.\ using supervised learning to reproduce expert labels). Instead, through DL we train sophisticated but imperfect agents, with differing sets of knowledge. For example, a bird classifier derived from an AudioSet embedding may have one type of expertise, while a raw waveform system trained from scratch has a different expertise. As the use of these systems becomes even more standardised, they take on the role of expert peers, with whom we consult and debate.
The move to active learning, which deserves more attention, cements this role by allowing DL agents to learn from criticism of their decisions.
Hence, DL does not displace the role of experts, nor even of crowdsourcing; future work in the field will integrate the benefits of all three \citep{Kitzes:2019}.
Secondly, open science is a vital component of progress. We have seen that the open publication of datasets, NN architectures, pretrained weights, and other source code has been crucial in the development of bioacoustic DL.
There is a move toward open sharing of data, but in bioacoustics this is incomplete \citep{Baker:2019}.
Sharing audio and metadata, and the standardisation of metadata, will help us to move far beyond the limitations of single datasets.

\subsection*{Maturing Topics? Architectures and Features}

Let us also briefly revise core topics within bioacoustic DL that are frequently discussed, but can be considered to be maturing, and thus not of high urgency.

The vast majority of the surveyed work uses spectrograms or mel spectrograms as the input data representation.
Although some authors raise the question of whether a species-customised spectrogram should be more appropriate than the human-derived mel spectrogram, for many tasks such alterations are unlikely to make a strong difference: as long as the spectrogram represents sufficient detail, and a DL algorithm can reasonably be trained, the empirical performance is likely to be similar across many different spectrogram representations.
Preprocessing such as noise reduction and PCEN is often found to be useful and will continue to be applied.
Methods based on raw waveforms, or adaptive front-ends such as SincNet or LEAF, are certainly of interest, and further exploration of these in bioacoustics is anticipated.
They may be particularly useful for tasks requiring fine-grained distinctions.

Commonly-used acoustic ``features'' in future are likely to include off-the-shelf deep embeddings, even more commonly than now.
Whether the input to those features is waveform or spectrogram will be irrelevant to users.
AudioSet and VGGish are the most commonly-used datasets for such pretraining;
note however that these cannot cover all bioacoustic needs---for example ultrasound---and so it seems likely that bioacoustics-specific embeddings will be useful in at least some niches.

CNNs have become dominant in many applications of DL, and this applies to bioacoustic DL too.
The more recent use of one-dimensional temporal convolutions (TCNs) is likely to continue, because of their simplicity and relative efficiency.
However, looking forward it is not in fact clear whether CNNs will retain their singular dominance.
In NLP and other domains, NN architectures based on ``attention'' (transformers/perceivers, discussed above) have displaced CNN as a basic architecture.
CNNs fit well with waveform and spectrogram data, and thus are likely to continue to contribute to NN architectures for sound, perhaps combined with transformer layers.
For example, \citet{Wolters:2021} propose to address sound event detection by using a CNN together with a perceiver network: their results imply that a perceiver is a good way to process variable-length spectrogram data into per-event summary representations.

A similar lesson applies to RNNs, except that RNNs have a more varied history of popularity. Recent CRNNs make good use of recurrent layers; but TCNs seem to threaten to displace them.
I suggest that although RNNs embody a very general idea about sequential data, they are a special case of more general \textit{computation with memory}. Transformers and other attention NNs show a different approach to allowing a computation to refer back to previous time steps flexibly. (See also the Legendre memory unit explored by \citet{Gupta::2021}.) All are special cases, and future work in DL may move more towards the general goal of differentiable neural computing (Graves et al 2018). The fluctuating popularity of recurrence and attention depends on their convenience and reusability as efficient modules in a DL architecture, and the future of DL-with-memory is likely to undergo many changes. Computational bioacoustics will continue to use these and integrate short- and long-term memory with other contextual data.

\subsection*{Learning Without Large Datasets}

Bioacoustics in general will benefit from the increasing open availability of data.
However, this does not dissipate the oft-studied issue of small data:
project-specific recognition tasks will continue to arise,
including high-resolution discrimination tasks, and tasks for which transfer learning is inappropriate (e.g.\ due to the risk of bias) \citep{Morfi:2021}.
Many approaches to dealing with small datasets have been surveyed in the preceding text;
important for future work is for these approaches to be integrated together, and for their advantages and disadvantages to be clarified.
As shown in data challenges such as BirdCLEF and DCASE, pre-training, embeddings, multi-task learning and data augmentation all offer low-risk methods for improved generalisation.

Few-shot learning is a recent topic of interest; it is not clear whether it will continue long-term to be a separate ``task'' or will integrate with wider approaches, but it reflects a common need in bioacoustic practice.
Active learning (AL) is also a paradigm of recent interest, and of high importance.
It moves beyond the basic non-interactive model of most machine learning, in which a fixed training set is the only information available to a classifier.
In AL, there is a human-machine interaction of multiple iterations, in which (some) predictions from a system are shown to a user for feedback, and the user's feedback about correct and mistaken identifications is fed into the next round of optimisation of the algorithm.
I identify AL as high importance because it offers a principled way to make the most efficient use of a person's time in labelling or otherwise interacting with a system \citep{Qian:2017}.
It can be a highly effective way to deal with large datasets, including domain shift and other issues.
It has been used in bioacoustic DL \citep{Steinfath::2021,Allen::2021} but is under-explored, in part because its interactive nature makes it slightly more complex to design an AL evaluation.
It may be that future work will use something akin to few-shot learning as the first step in an AL process.

A very different approach to reduce the dependence on large data is to create entirely simulated datasets that can be used for training.
This is referred to as \textit{sim2real} in DL, and its usefulness depends on whether it is feasible to create good simulations of the phenomena to be analysed.
It goes beyond data augmentation in generating new data points rather than modifying existing ones.
It may thus be able to generate higher diversity of training data, at a cost of lower realism.
One notable advantage of sim2real is that any confounds or biases in the training data can be directly controlled.
Simulated datasets have been explored in training DL detectors of marine sounds, perhaps because this class of signals can be modelled using chirp/impulse/sinusoidal synthesis \citep{glotin2017fast,Yang::2021,Li::2020}.
Simulation is also especially relevant for spatial sound scenes, since the spatial details of natural sound scenes are hard to annotate \citep{gao2020visualechoes,Simon::2021}.
Simulation, often involving composing soundscapes from a library of sound clips, has been found useful in urban and domestic sound analysis \citep{Salamon:2017,Turpault:2021}.
Such results imply that wider use in bioacoustic DL may be productive, even when simulation of the sound types in question is not perfect.

\subsection*{Equal Representation}

Deep learning systems are well-known to be powerful but with two important weaknesses:
(1) in most cases they must be treated as `black boxes' whose detailed behaviour in response to new data is an empirical question;
(2) they can carry a high risk of making biased decisions, which usually occurs because they faithfully reproduce biases in the training data \citep{Koenecke:2020}.
Our concern here is to create DL systems that can be a \textit{reliable} guide to animal vocalisations, especially if used to guide conservation interventions.
Hence we should ensure that the tools we create lead to an equal representation in terms of their sensitivity, error rates, etc.\ \citep{Hardt:2016}.

\citet{Baker:2019} point out that research output in bioacoustics is strongly biased: its taxonomic balance is unrepresentative of the audible animal kingdom, whether considered in terms of species diversity, biomass, or conservation importance. The same is true in the sub-field of DL applied to bioacoustics, both for datasets and research papers (see taxa listed above). Baker advocates for further attention to insect sound, and insects are recognised more broadly as under-studied \citep{Montgomery:2020}; \citet{Linke:2018} make a related case for freshwater species.

Equal representation (taxonomic, geographic, etc.) can be inspected in a dataset, and we should make further efforts to join forces and create more diverse open datasets, covering urban and remote locations, rich and poor countries.
\citet{Baker:2019} argue that the general field of bioacoustic research suffers from a lack of data deposition, with only 21\% of studied papers publishing acoustic recordings for others to use. Addressing this gap in open science practice may in fact be our most accessible route to better coverage in acoustic data.

Equal representation should also be evaluated in feature representations such as widely-used embeddings. The representational capacities of an embedding derive from the dataset, the NN architecture and the training regime, and any of these factors could introduce biases that represent some acoustic behaviours better than others.

Beyond equal representation, it may of course remain important to develop targeted methods, such as those targeted at rare species \citep{Znidersic:2020,Wood::2021}. Since rare occurrences are intrinsically difficult to create large datasets for, this is worthy of further study.
This review lists many methods that may help when rare species are of interest, but the best use of them is not yet resolved.
To give examples beyond the bioacoustic literature: \citet{Beery:2019} explore the use of synthetic examples for rarely-observed categories (in camera trap images); and \citet{Baumann:2020} consider frameworks for evaluating rare sound event detection.

\subsection*{Interfaces and Visualisation}

Many bioacoustic DL projects end with their outputs as custom Python scripts: this is good practice in computer science/DL, for reproducibility, but not immediately accessible to a broad community of zoologists/conservationists.
User interfaces (UIs) are a non-trivial component in bridging this gap.
Since the potential users of DL may wish to use it via R, Python, desktop apps, smartphone apps, or websites, there remains no clear consensus on what kinds of UI will be most appropriate for bioacoustic DL, besides the general wish to integrate with existing audio editing/annotation tools.
It seems likely that in future many algorithms will be available as installable packages or web APIs, and accessed variously through R/Python/desktop/etc as preferred.
Some existing work creates and even evaluates interfaces (discussed above, Workflow section), but more work on this is merited, including (a) research on efficient human-computer interaction for bioacoustics, and (b) visualisation tools making use of large-scale DL processing (cf.\ \citet{Kholgi:2018,Znidersic:2020,Phillips:2018}).

One domain in which user interaction is particularly important is active learning (AL), since it involves an iterative human-computer interaction. The machine learning components in AL can be developed without UI work, but interaction with sound data has idiosyncratic characteristics (temporal regions, spectrograms, simultaneously-occurring sounds) which suggest that productive bioacoustic AL will involve UI designs that specifically enhance this interaction.

Beyond human-computer interaction is animal-computer interaction, for example using robotic animal agents in behavioural studies \citep{Simon:2019,Slonina:2021}. These studies offer the prospect of new insights about animal behaviour, and they might use DL in future to provide sophisticated vocal interaction.

The most common formulation of DL tasks, via fixed sets of training data and evaluation data, become less relevant when considering active learning and other interactive situations.
There will need to be further consideration of the format, for example of data-driven challenges,
and potentially DL techniques such as reinforcement learning (not reviewed here since not used in the current literature) \citep{Te_ileanu_2017}.

\subsection*{Under-Explored Machine Learning Tasks}

The following tasks are known in the literature, but according to the present survey are not yet mature, and also worthy of further work because of their importance or generality.

\subsubsection*{Individual ID}

Automatically recognising discriminating between individual animals has been addressed by many studies in bioacoustics, whether for understanding animal communication or for monitoring/censusing animals \citep{Ptacek:2016,Vignal:2008,Searby:2004,Linhart:2019,Adi:2010,Fox:2008,Beecher:1989}.
Acoustic recognition of individuals can be a non-invasive replacement for survey techniques involving physical capture of individuals; it thus holds potential for improved monitoring with lower disturbance of wild populations.
Thus far DL has only rarely been applied to individual ID in acoustic surveying, though this will surely change \citep{Ntalampiras::2021,Nolasco::2021}.
Within-species acoustic differences between individuals are typically fine-scale differences, requiring finer distinctions than species distinctions.
This makes the task harder than species classification.

I suggest that these characteristics make the task of general-purpose automatic discrimination of individual animals, a useful focus for DL development.
A DL system that can address this task usefully is one that can make use of diverse fine acoustic distinctions.
Its inferences will be of use in ethology as well as in biodiversity monitoring.
Cross-species approaches and multi-task learning can help to bridge bioacoustic considerations across the various taxon groups commonly studied \citep{Nolasco::2021}.
A complete approach to individual recognition would also handle the open-set issue well, since novel individuals may often be encountered in the wild.
There are not many bioacoustic datasets labelled with individual ID, and increased open data sharing can help with this.

\subsubsection*{Sound Event Detection and Object Detection}

For many reasons it can be useful to create a detailed ``transcript'' of the sound events within a recording, going beyond basic classification.
As with individual ID, this more detailed analysis can feed into both ethological and biodiversity analyses; its development goes hand-in-hand with higher-resolution bioacoustic DL.

As described in the earlier discussion of detection, there are at least two main approaches to this in existing literature.
One version of SED (Figure \ref{fig:bdetect_paradigms}b) follows the same model as automatic music transcription or speaker diarisation in other domains, and uses similar DL architectures to solve the problem \citep{Mesaros::2021,Morfi:2018j,Morfi::2021}.
An alternative approach inherits directly from object detection based on bounding boxes in images (Figure \ref{fig:bdetect_paradigms}c). This fits well when data annotations are given as time-frequency bounding boxes drawn on spectrograms. Solutions typically adapt well-known image object detector architectures such as YOLO and Faster R-CNN, which are quite different from the architectures used in other tasks
\citep{Venkatesh::2021,Shrestha::2021,Zsebok::2019,Coffey::2019}.
These two approaches each have their advantages.
For example, frequency ranges in sound events can sometimes be useful information, but can sometimes be ill-defined/unnecessary, and not present in many datasets.
Future work in bioacoustic DL should take the best of each paradigm, perhaps with a unified approach that can be applied whether or not frequency bounds are included in the data about a sound event.

\subsubsection*{Spatial acoustics}

On a fine scale, the spatial arrangement of sound sources in a scene can be highly informative, for example in attributing calls to individuals and/or counting individuals correctly.
It can also be important for behavioural and evolutionary ecological analysis \citep{Jain:2011}.
Spatial location can be handled using multi-microphone arrays, including stereo or ambisonic microphones.
It is often analysed in terms of the direction-of-arrival (DoA) and/or range (distance) relative to the sensor.
Taken together, the DoA and the range imply the Cartesian location; but either of them can be useful on its own.

The standard approach to spatial analysis uses signal processing algorithms, even when the data are later to be classified using machine learning \citep{Kojima::2018}.
However, this may change. For example, \citet{Houegnigan::2017} train an MLP and \citet{VanKomen::2020} a CNN, to estimate the range (distance) of underwater synthetic sound events.
\citet{Yip:2019} perform a similar task using terrestrial recordings: using calibrated microphone recordings of two bird species, they obtain useful estimates of distance by deep learning regression from the sound level measurements.
In other domains of acoustics e.g.\ speech and urban sound, there is already a strong move to supplant signal processing with DL for spatial tasks 
\citep{Hammer_2021,Adavanne:2017b,Shimada:2021}.
Important to note is that these works usually deal with \textit{indoor} sound.
Indoor and outdoor environments have very different acoustic propagation effects, meaning that the generalisation to outdoor sound may not be trivial \citep{Traer:2016}.

Spatial inference can also be combined with SED (e.g.\ in the DCASE challenge ``sound event localisation and detection'' task or SELD), combining the two-step process (e.g.\ \citet{Kojima::2018}) into a single joint estimation task \citep{Shimada:2021}.

It is clear that many bioacoustic datasets and research questions will continue to be addressed in a spatially-agnostic fashion.
Although some spatial attributes such as distance can be estimated from single-channel recordings (as above), multi-channel audio is usually required for robust spatial inference.
Spatial acoustic considerations are quite different in terrestrial and marine sound, and more commonly considered in the latter.
However, the development of DL tasks such as distance estimation and SELD (in parallel to SED) could benefit bioacoustics generally, with local spatial information used more widely in analysis.

The discussion thus far does not address the broader geographic-scale distribution of populations, which statistical ecologists may estimate from observations.
Although machine observations will increasingly feed into such work, the use of DL in statistical ecology is outside the scope of this review (but cf.\ \citet{Kitzes:2019}).


\subsection*{Useful Integration of Outputs}

As DL becomes increasingly used in practice, there will inevitably be further work on integrating it into practical workflows, discussed earlier.
However, there are some gaps to be bridged, worthy of specific attention.

An important issue is the calibration of the outputs of automatic inference. \citet{Kitzes:2019} state the problem:
\begin{quote}
    ``We wish to specifically highlight one subtler challenge, however, which we believe is substantially hindering progress: the need for better approaches for dealing with uncertainty in these indirect observations. [...] First, machine learning classifiers must be specifically designed to return probabilistic, not binary, estimates of species occurrence in an image or recording. Second, statistical models must be designed to take this probabilistic classifier output as input data, instead of the more usual binary presence–absence data. The standard statistical models that are widely used in ecology and conservation, including generalized linear mixed models, generalized additive models and generalized estimating equations, are not designed for this type of input.'' \citep{Kitzes:2019}
\end{quote}
In fact, although many ML algorithms do output strict binary decisions, DL classifiers/detectors do not: they output numerical values between zero and one, which we can interpret as probabilities, or convert into binary decisions by thresholding.
However, the authors' first point does not disappear since
the outputs from DL systems are not always well-calibrated probabilities: 
they may under- or over-confident, depending on subtleties of how they have been trained (such as regularisation) \citep{Niculescu:2005}.
This does not present an issue when evaluating DL by standard metrics, but becomes clear when combining many automatic detections to form abundance estimates. DL outputs, interpreted as probabilities, may be under- or over-confident, or biased in favour of some categories and against others.
Measuring (mis)calibration is the first step, and postprocessing the outputs can help  \citep{Niculescu:2005}.
Evaluating systematic biases is also important: DL can be expected to exhibit higher sensitivity towards sounds well-represented in its training data, and this has been seen in practice \citep{Lostanlen::2018}.
Birdsong species classifiers are strongest for single-species recordings, and even with current DL they show reduced performance in denser soundscape recordings---an important concern given that much birdsong is heard in dense dawn choruses \citep{Joly::2019}.
Evaluating and improving upon these biases is vital.

The spatial reliability of detection is one particular facet of this.
For manual surveys, there is well-developed statistical methodology to measure how detection probability relates to the distance to the observer, and how this might vary with species and habitat type \citep{Johnston:2014}.
The same must be applied to automatic detectors. We have an advantage of reproducibility: we can assume that distance curves and calibration curves for a given DL algorithm, analysing audio from a given device model, will be largely consistent. Thus such measurements applied to a widely-used DL algorithm and recording device would be widely useful.
Some work does evaluate the performance of bioacoustic DL systems and how they degrade over distance \citep{Maegawa::2021,Lostanlen::2021vihar}.
This can be developed further, in both simulated and real acoustic environments.

Under the model of detection as binary classification, our observations are ``occupancy'' (presence/absence) measurements. These can be used to estimate population distributions, but are less informative than observed abundances of animals. Under the more detailed models of detection, we can recover individual calls/song bouts and then count them, though of course these do not directly reflect the number of animals unless we can use calling-rate information collected separately \citep{Stevenson:2015}.
Routes toward bridging this gap using DL include applying ``language models'' of vocal sequences and interactions; the use of spatial information to segregate calls per individual; and direct inference of animal abundance, skipping the intermediate step of call detections.
Counting and density estimation using DL has been explored for image data (e.g.\ \citet{Arteta:2016}), a kind of highly nonlinear regression task. Early work explores this for audio, using a CNN to predict the numbers of targeted bird/anuran species in a sound clip \citep{Dias::2021}.
\citet{Sethi:2020c} suggest that regression directly from deep acoustic embedding features to species relative density can work well, especially for common species with strong temporal occurrence patterns.

As DL tools become integrated into various workflows, the issue of standardised data exchange becomes salient.
Standards bodies such as TDWG provide guidance on formats for biodiversity data exchange, including those for machine observations.\footnote{\url{https://www.tdwg.org/}}
These standards are useful, but may require further development: for example, the probabilistic rather than binary output referenced by \citet{Kitzes:2019} needs to be usefully represented.
The attribution of observations to a specific algorithm (trained using specific datasets...) requires a refinement of the more conventional attribution metadata schemes used for named persons.
Such attribution can perhaps already be represented by standards such as the W3C Provenance Ontology, though such usage is not widespread.\footnote{\url{https://www.w3.org/TR/prov-o/}}

Taken together, these integration-related technical topics are important for closing the loop between bioacoustic monitoring, data repositories, policy and interventions.
They are thus salient for bringing bioacoustic DL into full service to help address the biodiversity crisis.

\subsection*{Behaviour and Multi-Agent Interactions}

Animal behaviour research (ethology) can certainly benefit from automatic detection of vocalisations, for intra- and inter-species vocal interactions and other behaviour.
This will increasingly make use of SED/SELD/object-detection to transcribe sound scenes.
Prior ethology works have used correlational analysis, Markov models and network analysis, though it is difficult to construct general-purpose data-driven models of vocal sequencing \citep{Kershenbaum:2014b,Stowell:2016}.
Deep learning offers the flexibility to model multi-agent sound scenes and interactions, with recent work including neural point process models that may offer new tools \citep{Xiao:2019,Chen:2020}.

Ethology is not the only reason to consider (vocal) behaviour in DL.
The modelling just mentioned is analogous to the so-called ``language model'' that is typically used in automatic speech recognition (ASR) technology: when applied to new sound recordings, it acts as a prior on the temporal structure of sound events, which helps to disambiguate among potential transcriptions \citep{OShaughnessy:2003}.
This structural prior is missing in most approaches to acoustic detection/classification, which often implies that each sound event is assumed to occur with conditional independence from others.
Note that language modelling in ASR considers only one voice.
A grand challenge in bioacoustic DL could be to construct DL ``language models'' that incorporate flexible, open-set, multi-agent models of vocal sequences and interactions;
and to integrate these with SED/SELD/object-detection methods for sound scene transcription.
Note that SED/SELD/object-detection paradigms will also need to be improved: for example the standard approach to SED is not only closed-set, but does not transcribe overlapping sound events within the same category \citep{Stowell:2015b}.
Analogies with natural sound scene parsing may help to design useful approaches \citep{Chait:2020}.

\subsection*{Low Impact}

When advocating for computational work that might be large-scale or widely deployed, we have a duty to consider the wider impacts of deploying such technology: carbon footprint, and resource usage (e.g. rare earth minerals and e-waste considerations of electronics). 
\citet{Lostanlen:2021} offer a very good summary of these considerations in bioacoustic monitoring hardware, as well as a novel proposition to develop batteryless bioacoustic devices.

For DL, impacts are incurred while training a NN, and while applying it in practice: their relative significance depends on whether training or inference will be run many times \citep{Henderson:2020}.
Happily, the power (and thus carbon emission) impacts of training DL can be reduced through some of the techniques that are already in favour for cross-task generalisation: using pretrained networks rather than starting training from random intialisation, and using pretrained embeddings as fixed feature transformations.
Data augmentation during training can \textit{increase} power usage by artificially increasing the training set size, but this can be offset if the required number of training epochs is reduced.
The increasing trend of NN architectures designed to have a low number of parameters or calculations (ResNet, MobileNet, EfficientNet) also helps to reduce the power intensity.

The question of running DL algorithms on-device brings further interesting resource tradeoffs.
Small devices may be highly efficient, and might neccessarily run small-footprint NNs.
(Note that a given algorithm may have differing footprint when run as a Python script on a general-purpose device, versus a low-level implementation for a custom device.)
Running DL on-device also offers the ability to reduce storage and/or communication overheads, by discarding irrelevant data at an early stage.
Alternatively, in many cases it may be more efficient to use fixed recording schedules and analyse data later in batches \citep{Dekkers:2022}.
The question becomes still more complex when considering networking options such as GSM/LoRa or star- versus mesh-networking.

Our domain has only just begun to spell out these factors coherently.
Smart bioacoustic monitoring has potential to provide rapid-response ecosystem monitoring, in support of nature-based solutions in climate change and biodiversity.
This motivates further development in low-impact bioacoustic DL paradigms.


\section*{Conclusions}

In computational bioacoustics, as in other fields, DL has enabled a leap in the performance of automatic systems.
Bioacoustics will continue to benefit from wider developments in DL, including methods adapted from image recognition, speech and general audio.
However, it is not merely a question of adopting techniques from neighbouring fields.
The roadmap presented here identifies topics meriting study within bioacoustics, arising from the specific characteristics of the data and questions we face.



\bibliography{bibliography_compbioac_deeplearning_2022,../../../refs.bib}

\end{document}